\documentclass[aps,showpacs,nofootinbib]{revtex4} 

\usepackage{amsmath}
\usepackage{graphicx}
\usepackage{dcolumn}
\usepackage{bm}
\usepackage{color}
\usepackage[normalem]{ulem}

\usepackage{color}

\begin{document}

\title{Maximum configuration principle for driven systems with arbitrary driving}

\author{Rudolf Hanel$^{1,4}$ and Stefan Thurner$^{1,2,3,4}$
}
\email{stefan.thurner@meduniwien.ac.at} 

\affiliation{
$^1$ Section for the Science of Complex Systems, CeMSIIS, Medical University of Vienna, 
Spitalgasse 23, A-1090, Vienna, Austria\\
$^2$ Santa Fe Institute, 1399 Hyde Park Road, Santa Fe, NM 87501, USA\\
$^3$ IIASA, Schlossplatz 1, 2361 Laxenburg, Austria \\
$^4$ Complexity Science Hub Vienna, Josefst{\"a}dterstrasse 39, A-1090 Vienna, Austria \\
}


\begin{abstract}
Depending on context, the term entropy is used for a thermodynamic quantity, a~measure of available choice, 
a quantity to measure information, or, in the context of statistical inference,   a maximum configuration predictor. 
For systems in equilibrium or processes without memory, the mathematical expression for these different concepts of entropy  
appears to be the so-called Boltzmann--Gibbs--Shannon entropy, $H$. 
For processes with memory, such as driven- or self-reinforcing-processes, this is no longer true: 
the different entropy concepts lead to distinct functionals that  generally differ from $H$. 
Here we focus on the maximum configuration entropy (that predicts empirical distribution functions)
in the context of driven dissipative systems.  
We develop the corresponding framework and derive the entropy functional 
that describes the distribution of observable states as a function of the details of the driving process. 
We do this for sample space reducing (SSR) processes,  which provide an analytically tractable model for driven dissipative systems 
with controllable driving.  
The fact that a consistent framework for a maximum configuration entropy exists for arbitrarily driven non-equilibrium systems  
opens the possibility of deriving a full statistical theory of driven dissipative systems of this kind.
This provides us with the technical means needed~to derive a thermodynamic theory of driven processes based on a statistical theory. 
{
 We discuss the Legendre structure for driven systems.           
 } 
\end{abstract}


\maketitle

\section{Introduction}

Entropy is a single word that refers to a number of very distinct concepts, a fact that has caused much confusion.  
These concepts cover areas as diverse as statistical mechanics, minimal code lengths, or the multiplicity of available choices in statistical inference. 
The distinction between the concepts underlying entropy is frequently not made properly. 
As long as we deal with equilibrium processes or processes with no memory (that can be modeled as i.i.d. sampling processes), 
formally it does not matter which entropy concept we use: they all lead to the same functional
\begin{equation}
H(p)=-k\sum_{i\in\Omega} p_i\log p_i \quad
\label{Hshannon} 
\end{equation}
where $k$ is a context- and unit-dependent positive constant, and $\Omega = \{1,2,\cdots,W\}$ is the \textit{sample space} 
containing $W$ distinct states $i=1,\cdots, W$ that occur (or are sampled) with relative frequencies $p_i$. 
In~statistical mechanics, $\Omega$ is the set of all accessible microstates; in information theory, it is an alphabet of 
$W$ letters that can be coded in a code-alphabet using $b$ code-letters (typically $0$ and $1$ for $b=2$). 
Setting  $k=1/\log(b)$ often indicates that $H$ is used as a measure of \emph{information production}. 
If all we know about a class of messages is the frequency $p_i$ of letters $i$, 
then $H$ is a sharp lower bound on the average number of code-letters per letter, needed for coding messages in the original alphabet. 
In~physics, $k$ is the Boltzmann constant, measuring entropy in units  of energy per degree Kelvin. 
We~refer to $H$ as the Boltzmann--Gibbs--Shannon entropy. 

However, this degeneracy of entropy concepts with respect to $H$ vanishes for processes with memory.
For systems or processes away from equilibrium, such as for processes with memory,  
the different entropy (production) concepts lead to distinct, non-degenerate entropy functionals. 
``Entropy,''  in a way, becomes context-sensitive: the functional form of entropy measures differ from one class of processes to another. 
The different concepts link structural knowledge about a given system with the functional form of its corresponding entropy measure.
We demonstrated this fact for self-reinforcing processes, P\'{o}lya urn processes in particular \cite{Ref:3entropies,Ref:book,Ref:PolyaEnt}, 
for driven dissipative systems that can be modeled with slowly driven \textit{sample space reducing} (SSR) processes \cite{Ref:BRSstaircase,Ref:SSR_exponents,Ref:SSR_Cascades,Ref:SSR_ElastColl} 
and for multinomial mixture processes \cite{Ref:3entropies}. 
Consequently, for non-equilibrium processes, we need to properly distinguish entropies depending on the context they are used for.
In {Reference} \cite{Ref:3entropies}, we distinguish three ({there is no reason} that there could not be more than three)  
entropy concepts: 
\begin{itemize}
\item  The concept of  \textit{maximum configuration} (MC) predictions defines entropy, $S_{\rm MC}$, 
as the logarithm of the multiplicity of possible equivalent choices and plays a central role in predicting 
empirical distribution functions of a system. It plays a role in statistical mechanics \cite{Ref:Boltzmann,Ref:Planck}  
but can be extended naturally beyond physics in the context of statistical inference  \cite{Ref:Jaynes}. 
\item  The concept of entropy in \textit{information theory} \cite{Ref:Shannon,Ref:Kraft,Ref:McMillan} 
can be identified as the \textit{information production rate} (IP) of information sources. The corresponding entropy, $S_{\rm IP}$, 
is an asymptotically sharp lower bound on the loss-free compression that can be achieved in the limit of infinitely long messages. 
\item The necessity of extensive variables (e.g., in thermodynamics) leads us to define an {\em extensive} entropy, $S_{\rm TD}$, through an 
asymptotic \textit{scaling expansion} that relates entropy with the phase-space volume growth  as the size of a system changes~\cite{Ref:Classification,Ref:ScaExpansion}. ({Consistent 
scaling of extensive variables is a prerequisite for thermodynamics to make sense.
We therefore call the corresponding entropy measure ``thermodynamic,'' $S_{\rm TD}$, 
despite the fact that all of the three mentioned concepts may be required to characterize 
the full thermodynamics of a non-equilibrium system. 
However, to~avoid confusion, we will use the term \textit{thermodynamic scaling} entropy 
whenever it is necessary to distinguish the notion from Clausius' thermodynamic entropy, 
$S_{\rm Clausius}$, for which, for reversible processes, $dS_{\rm Clausius}=dQ/T$, holds.}) 
\end{itemize}

Here we are interested in the description of driven statistical systems. 
The interest is to know the most likely probability distribution for finding the system in its possible states. 
Observables can then be computed. 
The maximum configuration entropy is a way to compute this probability distribution.
In~the following, we aim to understand entropies of 
driven dissipative systems that are relatively simple in the sense that they exhibit {\em stationary distributions}. 
In particular, we are interested in the functional form of their maximum configuration entropy, $S_{\rm MC}$.

Driven systems are composed from antagonistic processes: 
a \textit{driving process} that ``loads'' or charges the system (e.g., from low energy states to higher states)
and a \textit{relaxing} or  discharging process that brings the system towards  a ``ground state'' 
(equilibrium or sink state). ({Both processes by themselves may be non-ergodic. By coupling both processes, the system 
may---especially when it assumes steady states---regain ergodicity}.) 
The driving process induces a probabilistic ``current'' that {\em breaks detailed balance}. 
This creates a ``preferred direction'' in the dynamics of the system. 

In previous work \cite{Ref:BRSstaircase,Ref:SSR_exponents}, we discussed driven processes in the limit of extremely slow driving rates. 
We find simple functional expressions for maximum configuration entropies that depend on the probabilities $p_i$ only.
These denote the probability for the system to occupy state $i$.
The situation for general driving process---which constitutes the essence of this paper---is more involved. 
To characterize an arbitrary (stochastic) driving process, it is necessary to specify the probability with which the 
system that is in a given state $i$ at a given time $t$ is driven to a ``higher'' state $j$ at time $t+1$. 
If we think of states being associated with energy levels, that would mean $\epsilon_j>\epsilon_i$. 
We use $p^*_i$ to denote the {\em probability for the system to be ``driven''} when it is in state $i$. 
It is clear that entropies for processes that are driven in a specific way will depend on the driving rate that is given by $p_i^*$. 
As a consequence, the entropy for driven dissipative systems is very different from those 
that were discussed in the literature of generalized entropies. 
We will see that the MC entropy of driven dissipative systems is neither of trace form, nor does it depend on one single distribution function only. 
The latter property means that they violate the first Shannon--Khinchin axiom, see, e.g.,  \cite{Ref:Classification}. 
The adequate entropies of driven systems become measures that jointly describe 
the relaxation and the driving of the system; they depend on $p_i$ and the (state-dependent) $p_i^*$.

Despite these complications, we will see that the \textit{maximum configuration entropy} of driven systems remains 
a meaningful and useful concept for predicting distribution functions of driven systems. 
We~further show that the maximum entropy principle for driven systems still follows the mantra: 
\textit{maximize an entropy functional  $+$ constraints}. 
This functional corresponds (up to a sign) to the information divergence or relative entropy of the driven process. 
By identifying the constraint terms with cross-entropy, the known formula holds also within the driven context: 
\mbox{entropy $=$ cross entropy $-$ divergence}.

\section{The Sample Space of Driven Dissipative Systems}

Imagine a driven system. It is composed of a driving process  and a relaxation process. 
The~latter determines the dynamics of the system in  the case of no driving---the system relaxes 
towards an attractor-, sink-, or equilibrium-state. 
The driving process can be thought of as a potential that keeps the system away from equilibrium. 
The system dynamics is then described by the details of the driving  and relaxation processes.
The collection of states $i$ that a system can take is called the discrete {\em sample space},  $\Omega$. 
For simplicity, we also assume discrete time steps. ({All of the following can be formulated in continuous variables. 
A continuous sample space we call \textit{phase space}}.)
The dynamics of a system is characterized by the temporal succession of states: ${\rm initial \ state} \to i\to j\to k\cdots \to{\rm final \ state}$. 
If the system is stochastic, the transition probabilities, $p(j|i)$, represent the probability that the system will be found in state $j$ in the next step, 
given that it is in state $i$ now.  
States are characterized by certain properties  that allows them to be sorted.  
In physical systems, such properties could be energy levels, $\varepsilon_i$, that are associated to state $i$. 
A relaxing process is a sequence of states that leads from high to lower energy levels. 
In~other words, $p(j|i)\geq0$ if $\varepsilon_i>\varepsilon_j$, and $p(j|i)=0$  if $\varepsilon_i<\varepsilon_j$. 
This is what we call a {\em sample space reducing process},  a process that reduces its sample space over time \cite{Ref:BRSstaircase,Ref:SSR_exponents}. 
The current sample space (and its volume) depends on the current state $i$ the system is in. It is the set 
of states that can be reached from its current state $i$, $\Omega(i)=\{j|\varepsilon_j<\varepsilon_i\}$. 
If a system relaxes in a sequence $i\to j\to k\cdots$, then the sequence of sample spaces 
$\Omega(i)\supset \Omega(j)\supset\Omega(k)\cdots$ is nested. 
Generally, relaxation processes are sample space reducing~processes.

\begin{figure}
\begin{center}
		\includegraphics[width=0.4\columnwidth]{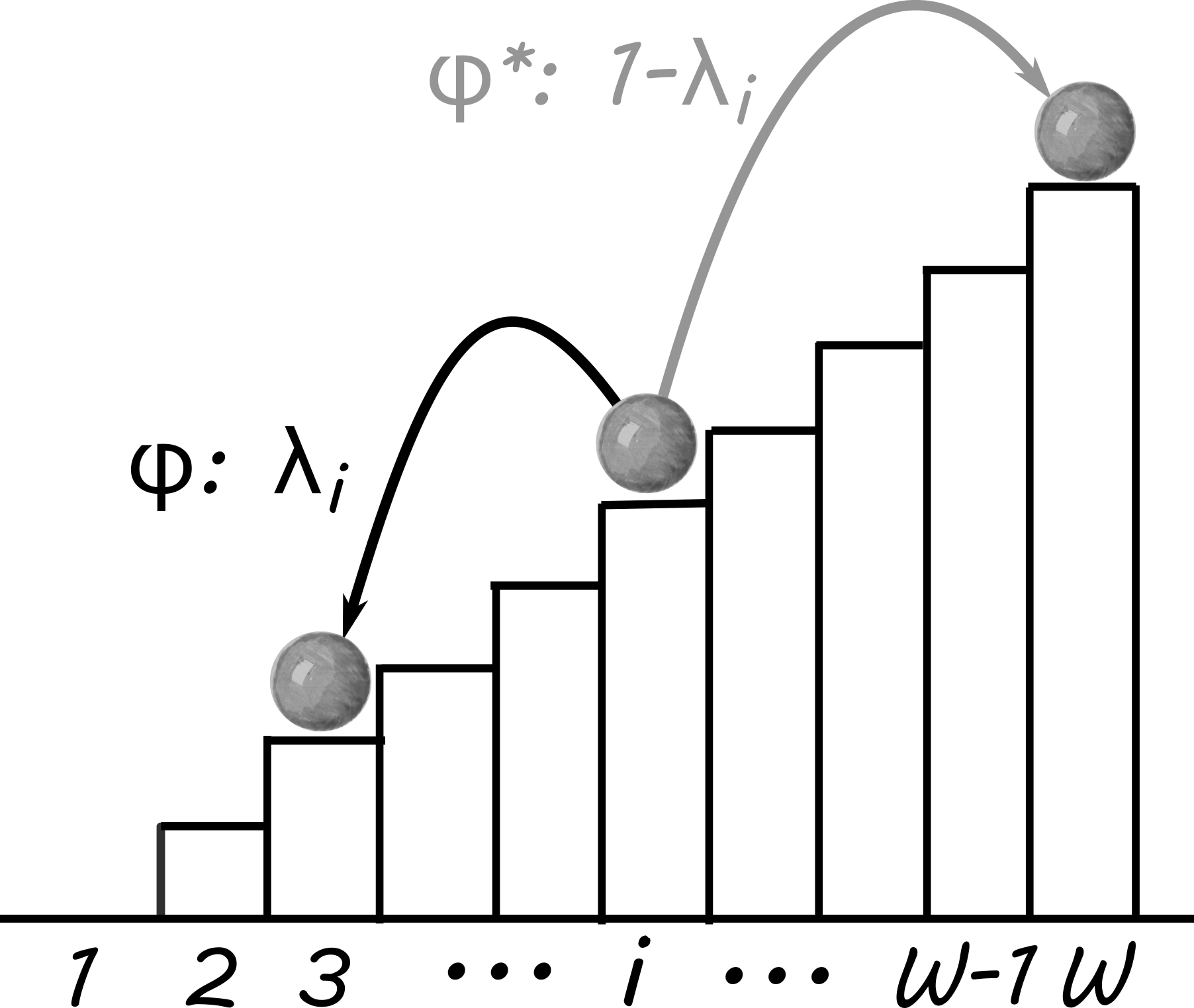}
\end{center}
	\caption{
	Illustration of a driven process as a combination of a relaxing process, $\phi$, and a driving process, $\phi^*$. 
	Without driving, the ball jumps randomly to any lower stair---it never jumps upward. It~follows a sample space reducing process until it hits 
	 the ground state $i=1$, upon which it is restarted. Restarting,  or driving, means that it becomes lifted to any higher state.
	 The system is driven after it is fully relaxed to the ground state. This scenario we call slow driving. 
	In a more general situation, at~any state $i$ there is a probability 
	 $1-\lambda_i$ that the process experiences a driving event in the next update.   
	 Intuitively,  with state-dependent driving, with probability $\lambda_i$, the process follows $\phi$ (black), and relaxes (can only sample states $j<i$, 
	 the ball moves downward),  or the process faces a driving event with probability $1-\lambda_i$, 
	 and follows the driving process, $\phi^*$ (grey).
	\label{Fig1}
	}
\end{figure}

A driving process  typically brings a system from ``low states'' to ``higher states''; or it brings low states to any other state. 
The sample space of the driving process can be the highest state, 
any state that is higher than the present state, or any state, i.e., the entire sample space, $\Omega_{\rm drive}=\Omega$. 
In other words, it can (but does not have to) depend on the current state. 
A driving event is any event that allows the system to change from a low state to a higher state. 
A schematic view of a  driven process is shown in Figure~\ref{Fig1}. 
This particular example consists of a relaxing (SSR) process, $\phi$,  and a driving process, $\phi^*$,  
that places the system always in the highest possible state.  
 
\subsection{Slowly Driven Sample Space Reducing Processes}\label{simplessr}

Let us briefly recapitulate the properties of sample space reducing processes with a simple example. 
Given a sample space with $W$ states, $\Omega={1,2,\cdots, W}$, 
imagine that the process samples state $x_1\in\Omega$. Then, in the next time step, the process can only sample from 
$\Omega(x_1)={1,2,\cdots, x_1-1}\subset \Omega$. 
First assume (without loss of generality~\cite{Ref:SSR_exponents}) that all of the states in $\Omega(x_1)$ are sampled with the same probability. 
With every sample $x_n$, the process is left with a smaller sample space, similar to a ball bouncing down a staircase, 
which can only sample stairs below its current position (see Figure  \ref{Fig1}). 
Once the process reaches the lowest possible state, $x_m=1$ at some point in the process (think of time), $m$, 
then the sample space of the process becomes empty and the process stops at the sink state. 
The process can only continue if the sample space reducing process is restarted with a driving event that 
will allow it to sample from larger sample spaces again.

The simplest way to specify the driving process in this example  is to allow the process to sample from the entire sample space $\Omega$, 
once it has arrived at the sink state, $i=1$. This is done by setting $\Omega(1) = \Omega$. 
In this case, the driving process becomes active after the system is completely relaxed. We~call that the {\em slow driving} limit. 
Restarting the relaxing sample space reducing process produces an ongoing Markov process, 
and a stationary visiting distribution function of the states $i$, $p_i$. 
However, detailed balance is explicitly violated in driven processes. 

For the slowly driven process, the transition probabilities are given by 
\begin{equation}
p(i|j)=\Theta(j-i) \frac{q_i} {Q_{j-1}}+\delta_{j1}  q_i \quad
\label{transprob1} 
\end{equation}
where $\Theta(x)$ is the Heaviside function, $\delta_{ij}$ is the Kronecker delta, and
$Q_j=\sum_{i=1}^j q_i$ is the cumulative distribution of weights (or prior probabilities, {which determine how likely the process starts in state $i$}) $q$.
Let $p=(p_1,\cdots,p_W)$ be the visiting distribution of states, i.e., $p_i$ 
is the relative frequency of state $i$ being  visited by the driven process, then 
the equation $p_j=\sum_{i=1}^W p(j|i)p_i$ holds and 
\begin{equation}
	p_i=p_1\frac{q_i}{Q_{i}}\quad.
\label{zipf1}
\end{equation}

For the choice $q_i=1/W$ (equidistributed prior probabilities),  the transition probabilities read
$p(i|j)=1/(j-1)$, and 
we immediately obtain
\begin{equation}
	p_i \sim \frac{1}{i}\quad, 
\end{equation}
i.e., Zipf's law.
Remarkably, also for almost any other choice of $q_i$, the same result is obtained, meaning that Zipf's law 
in the probability of state visits is an attractor of slowly driven systems, regardless of the details of the relaxing process. 
For details, see \cite{Ref:BRSstaircase,Ref:SSR_exponents, Ref:SSR_Cascades}.
A physical example is given in \cite{Ref:SSR_ElastColl}, where the kinetic energy of a projectile passing through 
a box filled with balls---with which it collides elastically---reduces as an sample space reducing  process. 
Zipf's law is found in the occurrence frequencies of projectiles with specific kinetic energies.

\subsection{Driven Sample Space Reducing Processes with Arbitrary Driving}\label{drivenssr}

Driven sample space reducing processes on the states $i=1,\cdots,W$ can be characterized as a mixture of the relaxing process 
$\phi$, and a (usually stochastic) driving process, $\phi^*$.
As before, let us assume (without loss of generality) that the driving process $\phi^*$  
samples among all possible states $i\in\Omega$ with probability $q_i$.  
In contrast to the slow driving case,  a driving event can now occur at any time, 
not only after the system reaches its ground state, $i=1$.
Symbolically, the combination of driving and relaxing processes reads $\phi_\lambda=\lambda \phi+(1-\lambda) \phi^*$, where 
$\lambda$ is the probability that the process follows $\phi$ (relaxes), and $1-\lambda$ is the 
probability for a driving event. 
In the ground state, $i=1$, $\phi$, and $\phi^*$ behave identically. 
In the general case, $\lambda$ may depend on the state $i$, i.e., $\lambda=(\lambda_1,\cdots,\lambda_W)$, and~$1-\lambda_i$ becomes  
 a state-dependent driving rate. 
The transition probability from state $j$ to state $i$ is
 \begin{equation}
p_\lambda(i|j)=q_i\left(\frac{\Theta(j-i)\lambda_j}{Q_{j-1}}+1-\Theta(j-1)\lambda_j\right)\quad.
\label{condprob}
\end{equation}  
Note that the value of $\lambda_1$ does not effect the transition probabilities of the process.
Technically, we set $\lambda_1=0$.
Using $p_i=\sum_{j=1}^Wp(i|j)p_j$, one finds the exact asymptotic steady state solution: 
\begin{equation}
\frac{p_i}{p_1}=\frac{q_i}{q_1}\prod_{j=2}^i\left(1+\lambda_j\frac{q_{j}}{Q_{j-1}}\right)^{-1} \quad.
\label{eigvsolR}
\end{equation}  
For the case where all $\lambda_j=0$, we get $p_i=q_i$, as expected for processes that are dominated by the driving process.
For all $\lambda_j=1$  (except for $\lambda_1$), we recover the slow driving limit of Equation~(\ref{zipf1}).
For the equidistributed prior probabilities, $q_i=1/W$, the following relation between the state-dependent driving rate, $1-\lambda(x)$, and the 
(stationary) distribution function exists
\begin{equation}
\lambda (x) = -x \frac{d}{dx} \log p(x) \quad.
\label{eigvsolSx}
\end{equation}  
Here, $x$ labels the states in a continuous way. For details,  see \cite{Ref:SSR_ElastColl}.

\section{The Maximum Configuration Entropy of Driven Processes}

For stationary systems, the probabilities for state visits can be derived with the use of the maximum entropy principle. 
We discuss the implications of driving a process for the underlying MC entropy functional. 
Let $k_i$ count how often a particular state $i=1,2,\cdots,W$ has occurred in $N$ sampling steps (think of a sequence of $N$ samples). 
$k=(k_1,\cdots,k_W)$ is the histogram of the process, and $p_i=k_i/N$ are the relative frequencies.
 
To find the maximally likely distribution (maximum configuration) of a process, one considers the probability 
to observe the specific histogram, $k$, after $N$ samples (steps), $P(k|\theta)=M(k)G(k|\theta)$.  
Here $M(k)$ is the multiplicity of the process, i.e. the number
of sequences $x$ that have the histogram $k$.  $G(k|\theta)$ is the probability of sampling one particular sequence with histogram $k$, and $theta$ parameterizes the process.
Taking~logarithms on both sides of $P(k|\theta)=M(k)G(k|\theta)$, we identify 
\begin{equation}
		\underbrace{\frac1R \log P(k|q)}_{\rm - relative \ entropy} =	{  \underbrace{  \frac1R\log M(k)}_{{\rm MC \ entropy} \ S_{MC}} }
		+ { \underbrace{  \frac1R \log G(k|q) }_{\rm - cross-entropy} } \quad. 
		\label{SMfactor}
\end{equation}
Here we scaled both sides by a factor, $R(N)$, that represents the number of degrees of freedom in the system.
To arrive at relative frequencies, we use $p=k/N$. 
Given that $N$ is sufficiently large, $p$~approaches the most likely distribution function. 

\subsection{Slowly Driven Processes}

Driving a process slowly means to restart it with a rate that is lower (or of the same order) than the typical relaxation time. 
In the above context of the sample space reducing process, it is restarted once the process has reached its ground state at $i=1$. 
The computation of the MC entropy, together with its associated distribution functions, 
for slowly driven systems has been done in \cite{Ref:3entropies, Ref:book}. 
We include it here for completeness. 
For simplicity, we present the following derivation in the framework of sequences. They hold more generally for 
other processes and systems. 

Let $x=(x_1,x_2,\cdots,x_N)$ be a sequence that records the visits  
to the states $i$. $\lambda=1$ means that the process is slowly driven and restarts only in state $i=1$.
To compute the multiplicity, $M$, note that we can decompose any sampled SSR sequence 
$x=(x_1,x_2,\cdots,x_N)$ into a concatenation of shorter sequences $x^{r}$ such that $x=x^{1}x^{2}\cdots x^{R}$. 
Each $x^r$ starts directly after a driving event restarted the process, and ends when the process halts at $i=1$,
where it awaits the next driving event. We call $x^r$ one ``run'' of the SSR process.
Since every run ends in the ground state $i=1$, we can be sure that, for the slowly driven process, we find $R=k_1$: number of runs is equal to the number of times we observe state $1$.
To determine  $M$ and the probability $G$, we represent the sequence $x$ in a table, 
where we arrange the runs $x^r$ in $W$ columns and $k_1$ rows, such that the entry in the $r$th row and $i$th
column contains the symbol ``$*$'' if run $r$ visited state $i$, and the symbol ``$-$'' otherwise.
\begin{equation}
	\begin{array}{c|c|c|c| c |c|c}
	 r \times i  & W & W-1 & W-2 & \cdots & 2 & 1 \\
	\hline
	 1 & - & * & - & \cdots & * & * \\
	 2 & * & - & * & \cdots & - & * \\
	 3 & * & * & - & \cdots & - & * \\
	 \vdots & \vdots & \vdots & \vdots &  & \vdots & \vdots \\
	 R-2 & - & * & * & \cdots & - & * \\
	 R-1 & - & * & - & \cdots & * & * \\
	 R & - & * & * & \cdots & - & * \\
	\hline
	  & k_W & k_{W-1} & k_{W-2} & \cdots & k_2 & k_1 \\
	\end{array}
	\label{table1}
\end{equation}

Recognizing that each column $i$ is $k_1=R$ entries long and contains $k_i\leq k_1$ items $*$, 
it~follows that without changing a histogram $k$ we can rearrange the items $*$ in column $i$ in   
\mbox{${k_1 \choose{k_i}}=k_1!/k_i!/(k_1-k_i)!$} ways.   
$M$ therefore is given by the product of binomial factors  
\begin{equation}
	M(k)=\prod_{i=2}^W{k_1 \choose{k_i}}\quad.
\label{scmult}
\end{equation} 

In other words, $M(k)$ counts all possible sequences $x$ with histogram $k$.
Using Stirling's approximation, the MC entropy, $S_{\rm MC}(p)= \frac1N\log M(Np)$, is 
\begin{equation}
	S_{\rm MC}(p)=-\sum_{i=2}^W \left[ p_i\log\left(\frac{p_i}{p_1}\right)+(p_1-p_i)\log\left(1-\frac{p_i}{p_1}\right)\right]\quad.
\label{scent}
\end{equation} 

Note that for computing the MC entropy of the slowly driven process 
one only needs to know that the process gets restarted at $i=1$.
At no point did we need to know the transition probabilities from Equation~(\ref{transprob1}).

This has an important implication: the entropy of the slowly driven process does not depend on 
how a process relaxes exactly, it only matters that the process {\em does} relax somehow ({replacing the  transition probabilities Equation~(\ref{transprob1})
with other transition probabilities that respect the sample space reducing structure of the processes and restarting 
 at $i=1$
only changes the expression for the sequence probabilities, $G$}) and is restarted in state $1$. 

Using Equation~(\ref{transprob1}), the probability, $G$, for sampling a particular $x$ can be computed in a similar way.
Each visit to any state above the ground state, $i>1$, in the sequence $x$ contributes to the probability of the next 
visit to a state $j<i$ with a factor $Q_{i-1}$,
independently from the state $j$ that will be sampled next ({only if $i=1$   do we not obtain such a renormalizing factor,  
since the process restarts with all possible states $i\in\Omega$ as valid successor states of state $1$, with prior probability $q_i$.}). 
As a consequence, one obtains
\begin{equation}
	G(k|q,N)=\prod_{i=1}^W q_i^{k_i} \prod_{j=2}^W Q_{i-1}^{-k_i}\quad.
\label{scxprob}
\end{equation} 

The cross-entropy $S_{\rm cross}(p|q)=-\frac1N\log G(k|q,N)$ is given by
\begin{equation}
	S_{\rm cross}(p|q)=-\sum_{i=1}^W p_i\log q_i + \sum_{i=2}^W p_i\log Q_{i-1}\quad,
\label{sccross}
\end{equation} 
and the maximum entropy functional of the slowly driven staircase process is given by 
\mbox{$\psi=S_{\rm MC} -S_{\rm cross}$}.
Note that $-\psi(p|\theta)$ plays the role of a divergence,
generalizing the Kullback--Leibler divergence. 

To obtain the maximum configuration $\hat k$ or $\hat p=\hat k/N$, we can solve the maximum entropy principle by maximizing
$\psi$ with respect to $p$ under the constraint $\sum_{i=1}^W p_i=1$. 
As a result, one obtains, $p_i=p_1\frac{q_i}{Q_{i}}$, which is exactly the solution that we obtain  
in Equation~(\ref{zipf1}). 

Note that MC entropy is not of trace form and the driving process ``entangles'' the ground state $i=1$ with every other state.
This entropy violates all but one Shannon--Khinchin axiom. For the uniform distribution $p_i=1/W$, one obtains  
$S_{\rm MC}(1/W,\cdots,1/W)=0$. 

\subsection{Arbitrarily Driven Processes}

For exploring relaxing systems with arbitrary drivers, again, we have to compute the probability of finding the histogram, 
$P$, by following the generic construction rule for the MC divergence, entropy, and cross entropy (see \cite{Ref:HTMGM3}). 
We will see that, for arbitrarily driven processes, $P$  factorizes again into a well defined multiplicity, $M$,  
and a sequence probability, $G$, just as for slowly driven processes.

To find $P$, we analyze the properties of a sampled sequence $x$.
First note  that, by mixing $\phi$ and $\phi^*$,   we now have to determine at each step,  
whether the process  follows $\phi$ (relaxes) or $\phi^*$ (is reloaded). 
To~model this stochastically, assume binary experiments, using, e.g., a coin $\omega_n$ for every sample $x_n$.
If~the coin samples $\omega_n=0$ with probability $\lambda_{x_n}$, the process relaxes and follows $\phi$. 
If the coin samples $\omega_n=1$ with probability $1-\lambda_{x_n}$, the process follows $\phi^*$, and a driving event occurs. 
Let's call any instance, $x_n$, a {\em starting point of a run}, if   (i) $n=1$,   (ii) $\omega_n=0$, or (iii) $x_{n-1}=1$. 
We call an instance, $x_n$, an {\em end point of the run}, if   (i) $n=N$,   (ii) $\omega_{n}=1$, or (iii) $x_{n}=1$.
Except for $n=1$ and $n=N$, all starting points directly follow an endpoint, 
and we can decompose the sequence $x$ into $R$ runs, $x^r$, as before, such that $x=x^1x^2\cdots x^R$.
Each run, $x^r=(x_{\underline r},\cdots,x_{\overline r})$, begins at a starting point $x_{\underline r}$, 
and ends in an endpoint, $x_{\overline r}$. 
Note that $\underline 1=1$, $\overline r + 1=\underline{r+1}$, and $\overline R=N$.
The average length of the sequences $x^r$ is given by $\bar \ell=N/R$.

\subsubsection{MC Entropy of Arbitrarily Driven Processes}

Let $x^*=(x_{\overline 1},x_{\overline 2},\cdots,x_{\overline R})$ be the subsequence of $x$ containing all endpoints of runs $x^r$.
\mbox{$k^*=(k^*_1,\cdots,k^*_W)$}  is the histogram of $x^*$, with $\sum_i k^*_i=R$.   
We can now determine the number $M(k,k^*)$ of all sequences of runs $x=(x^1x^2\cdots x^R)$ with histograms $k$ and $k^*$. 
Similar to the case of slow driving, this can be done by arranging runs in a table of $R$ rows and $W$ columns
\begin{equation}
	\begin{array}{c|c|c|c| c |c|c|c}
	 r \times i  & W & W-1 & W-2 & \cdots & 2 & 1 & \\
	\hline
	 1 & * & - & - & \cdots & \bullet & - & l(x^1)\\
	 2 & - & * & * & \cdots & \bullet & - & l(x^2)\\
	 3 & * & - & \bullet & \cdots & - & - & l(x^3)\\
	 \vdots &  & \vdots &  &  &  & \vdots & \vdots\\
	 R-2 & - & * & * & \cdots & - & \bullet & l(x^{R-2})\\
	 R-1 & - & - & - & \cdots & \bullet & - & l(x^{R-1})\\
	 R & - & - & * & \cdots & * & \bullet & l(x^R)\\
	\hline
	  & k_W & k_{W-1} & k_{W-2} & \cdots & k_2 & k_1 & \\
	\end{array}.
	\label{table2}
\end{equation}

\begin{figure}
\begin{center}
		\includegraphics[width=0.4\columnwidth]{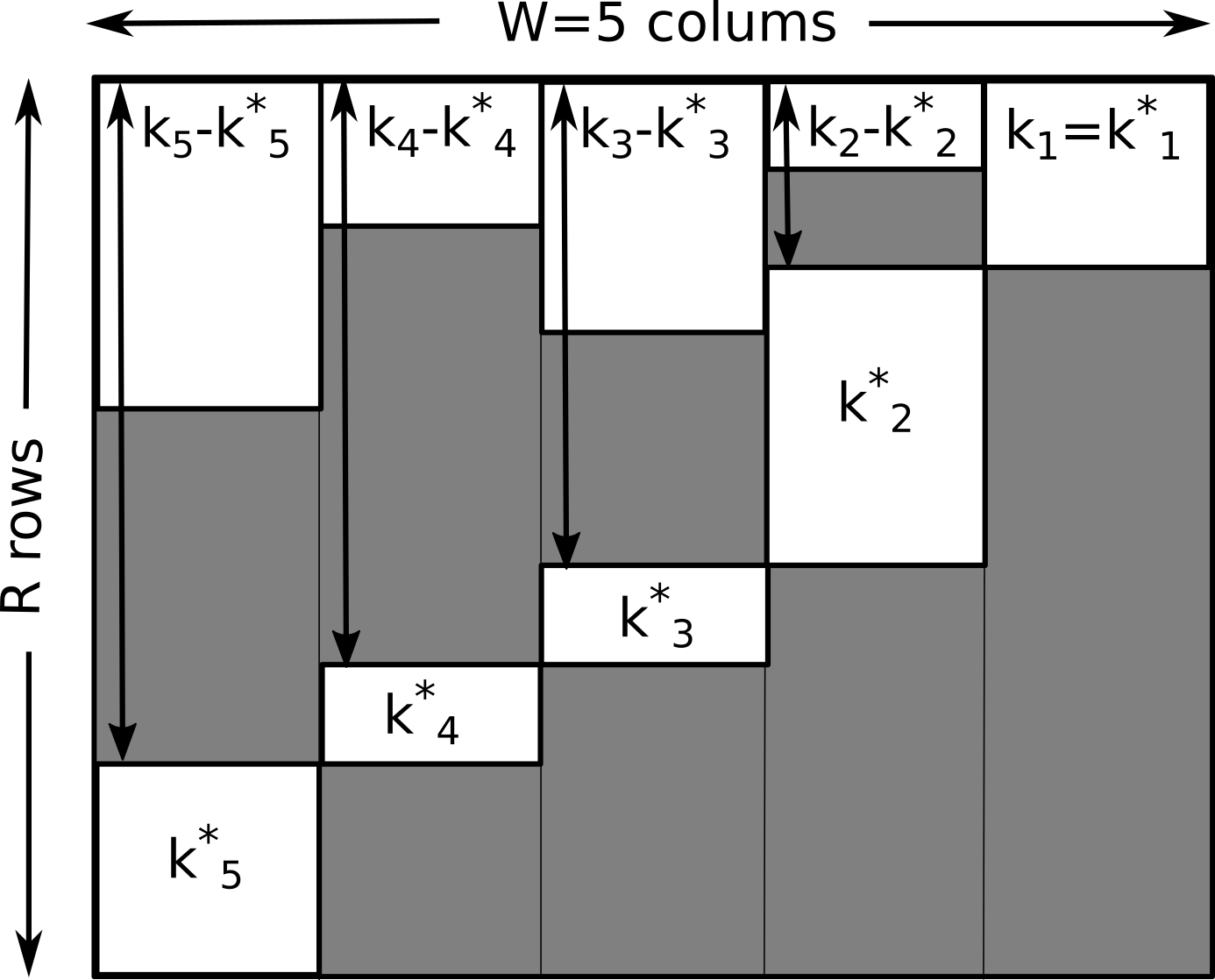}
\end{center}
	\caption{Demonstration of how the ordering of subsequences of a sequence $x$ (in increasing order of their endpoints)
	leads to a diagram that allows us to identify the combinatorial factors that determine the multiplicity, $M$, of sequences with  
	particular histograms $k$ and $k^*$.
	\label{Fig2}
	}
\end{figure}

We mark endpoints with the symbol ``$\bullet$'' (one per row), other sequence elements with ``$*$'', and the rest with ``$-$''.
Having fixed $k^*$, we can distribute the elements of $x^*$ in ${R \choose{k^*}}=R!/\prod_i k^*_i!$ different ways over $R$ rows. 
Given this fact, we can reorder the rows of the table in increasing order of their endpoints $x_{\overline r}$ (compare with Figure~\ref{Fig2}). Obviously  $k_1=k^*_1$, since all visits to the ground state are endpoints. 
We note further that for a state $i>1$, we can redistribute $k_j-k^*_j$ items $*$ over $\sum_{s=1}^{j-1} k^*_s$
available positions. We~can now simply read off the multiplicity from the ordered table
\begin{equation}
	M(k^*,k)={R \choose{k^*}}\prod_{j=2}^{W}{\sum_{s=1}^{j-1} k^*_s\choose{k_j-k^*_j}}\quad
	\label{multiplicity}
\end{equation}
where the product over the states $j$ are contributions of binomial factors counting the number of ways $k_i-k^*_i$ 
states can be distributed over $\sum_{s=1}^{j-1} k^*_s$ possible positions.

The multiplicity, $M$, does not explicitly depend on how exactly the system is driven. 
$M$ contains no traces of the driving rates $1-\lambda_i$, nor any dependencies on transition probabilities, 
meaning no 
dependencies on prior weights, $q_i$. 
This independence, however, is paid for with the dependence of $M$ on a second histogram, $k^*$.  $M$ depends only on the {\em combinatorial relations} of the relaxing (sample space reducing) process with the driving process. 
$M$ describes the multiplicity, i.e., the {\em a priori} chances of jointly observing $k$ and $k^*$ together. 
Note that $k^*_i$ is the empirically observed number of driving events observed in state $i$, 
$k_i$ is the empirical number of visits of the process to state $i$, while  
$1-\lambda_i$ is the probability of a driving event to occur when the system is in state $i$. 

The ``joint'' MC entropy  of the arbitrarily driven process, $\phi_\lambda$, is the logarithm of $M$ times a scaling factor. 
If we want an entropy per driving event, the factor is $1/R=\bar\ell/N$, and we obtain  
\begin{equation}
\begin{array}{l}
	S_{\rm MC}(p^*,p,\bar \ell)=\frac1R\log M(k^*,k)
	=H(p^*)+\sum_{j=2}^W\left(\sum_{s=1}^{j-1}p_s^*\right)H_2\left(\frac{\bar \ell p_j-p_j^*}{\sum_{z=1}^{j-1}p_z^*}\right)\quad,
\end{array}
	\label{dSSRMCEnt}
\end{equation} 
which depends on the average length of the runs, $\bar\ell=N/R$.
The parameter $\bar \ell$ corresponds to the \textit{average free relaxation time} 
of the process. Note that we can define the \textit{average driving rate} by $1-\bar \lambda=1/\bar\ell$,
which~is nothing but the ratio $R/N$, i.e., 
 the average number of driving events per process step.
The~MC entropy also depends on $p^*=k^*/R$ and $p=k/N$, the distributions of driving events and state visits, respectively. 
$H_2(x)=-x\log(x)-(1-x)\log(1-x)$ is the Shannon entropy of a binary process. 
From~the fact that, by definition, $k_1= k^*_1$, we obtain the constraint
\begin{equation}
	p_1^*= \bar \ell\ p_1 \quad.
\label{scaconstr}
\end{equation}

\subsubsection{Cross-Entropy and Divergence of Arbitrarily Driven Processes}

Similarly, we can collect the contributions to the sequence probability, $G$.
Every time the sequence $x$ follows the relaxing (or sample space reducing) process, we obtain a factor 
$\lambda_i q_i/Q_{i-1}$, for some $i=x_n$, and a factor $(1-\lambda_i)q_i$ if the 
sequence reaches an endpoint, $i=x_n$, of a run $x^r$.
As a consequence, the~probability of observing a sequence $x$ with histograms $k$ and $k^*$ is given by 
\begin{equation}
G(k^*,k|q,\lambda)=q_1^{k^*_1}\prod_{i=2}^W \left(\frac{\lambda_iq_i}{Q_{i-1}}\right)^{k_i-k^*_i}\prod_{j=2}^W\left(\left(1-\lambda_j\right)q_j\right)^{k^*_j} 
\left(\frac{\lambda_j }{Q_{j-1}}\right)^{k_j-k^*_j} \quad.
\label{dSSRG} 
\end{equation} 
Note that $N\geq R\geq\max(k_1\cdots,k_W)$. 
The joint probability of observing $k^*$ and $k$ together becomes
\begin{equation}
	P_\lambda(k^*,k|q,\lambda)=M(k^*,k)G(k^*,k|q,\lambda)\quad.
\end{equation} 

This shows that the mixing of relaxing and driving processes, $\phi$ and $\phi^*$, 
leads to a joint probability that depends on $k$ and $k^*$. Noting that we can define $\lambda_1=0$, the cross entropy construction yields 
\begin{equation}
\begin{array}{l}
S_{\rm cross}(p^*,p,\bar \ell|q,\lambda)=-\frac1R\log G(k^*,k|q,\lambda)\\
{
\quad=- \sum_{j=1}^W   p^*_j\left(\log (1-\lambda_j) + \log q_j\right) 
}
-\sum_{j=2}^W \left(\bar\ell p_j- p^*_j\right)\log\left(\frac{\lambda_j q_j}{Q_{j-1}}\right)\quad
\end{array}.
\label{crossent}
\end{equation} 

The MC functional, $\psi=\frac{1}{R}\log P$, is   
$\psi(p^*,p,\bar \ell|q,\lambda)= S_{\rm MC}(p^*,p,\bar \ell)-S_{\rm cross}(p^*,p,\bar \ell|q,\lambda)$.
$-\psi$ can be interpreted as the \textit{information divergence} or \textit{relative entropy} of the process.
Maximizing $\psi$ with respect to $p$, $p^*$, and $\bar \ell$, under the constraints
$\sum_i p_i^*=1$, $\sum_i p_i=1$, and $p^*_1=\bar\ell\ p_1$ (compare with Equation~(\ref{scaconstr})),
yields the maximum configuration $(\hat p, \hat p^*, \hat \ell)$.
The maximization of the MC functional,
\begin{equation}
	\psi-\alpha\left(\sum_ip_i-1\right)-\alpha^*\left(\sum_i p^*_i-1\right)-\gamma\left(\bar \ell p_1-p_1^*\right)\quad,
\label{MCfunc} 
\end{equation} 
with respect to $p$, $p^*$, and 
{
$\bar \ell$,
}
is the \textit{maximum entropy principle} of a driven processes. 
$\alpha$, $\alpha^*$, and $\gamma$ are Lagrangian multipliers.

By noting that derivatives of the MC functional Equation~(\ref{MCfunc}) behave differently with respect to $p_1$ and $p_1^*$, when  compared with how they behave with respect to $p_i$ and $p^*_i$ ($i>1$), one finds that $\alpha=\gamma=0$, and, using the short hand for $\mu_i=\lambda_iq_i/Q_{i-1}$, 
one obtains the formulas 
\begin{equation}
\begin{array}{lcl}
p_i&=&\frac1{\bar \ell}\left[p^*_i +\left(\sum_{s=1}^{i-1}p^*_s\right)\frac{\mu_i}{1+\mu_i}\right]\\
p^*_1&=&\frac{1}{Z^*}\prod_{s=2}^{W}\frac{\mu_s}{1+\mu_s}\\
p^*_i&=&\frac{(1-\lambda_i)q_i}{1+\mu_i}\frac{p^*_1}{q_1}\prod_{s=2}^{i-1}\frac{1}{1+\mu_s}\,\quad i>1\quad
\end{array}.
\end{equation}

These allow us to compute the solution for $p_i$ recursively. The result is 
exactly the prediction  from the eigenvector equation in Equation~(\ref{eigvsolR}).
This fact shows that, indeed, the maximum entropy principle yields the expected distributions. 
However, we learn much more here, since by construction, $\exp(-\psi(p,p^*,\bar \ell))$, is the  
probability measure of the process. 
This allows us to study fluctuations of $p$ and $p^*$ within a common framework. 
This framework opens the possibility of extending fluctuation theorems---which are of interest in statistical physics (see, e.g., \cite{Jarzinski})---to systems explicitly driven away from equilibrium.
We can immediately predict parameters of driven processes, such as  the 
“average free relaxation time” of the process, $\bar\ell=p^*_1/p_1$.

\section{Legendre Structure and Thermodynamic Interpretation}

In the standard equilibrium MC principle, one maximizes  the functional 
$H(p)-\alpha' \sum_i p_i-\beta' \sum_i p_i\varepsilon_i$, with respect to $p$, $\alpha'$, and $\beta'$. $\alpha'$ and $\beta'$ are Lagrange multipliers for the normalization 
constraint, $\sum_i p_i=1$, and (e.g., in a physics context) the  ``energy'' constraint, $\sum_i p_i\varepsilon_i=U$,
where $U$ represents the ``internal energy.''
$\varepsilon_i$ are ``energy'' states and at the extremal value, $\beta'=\hat\beta$ is the inverse temperature.
One can also see $\beta'=\hat \beta=\beta$ as a parameter $\beta$ of the extremal principle, and think of $U$ as a function of $\beta\varepsilon$.

The  classic MC principle can be obtained from 
{maximizing the functional \mbox{$\psi(p|q)=\frac{1}{N}\log P_{\rm multinomial}(Np|q)$} (in the large $N$ limit), 
under the constraint, $\sum_i p_i=1$. Asymptotically, one obtains
$\psi=H(p)-H(p|q)$
}
where $H(p|q)=-\sum_i p_i\log q_i$ is the equilibrium cross-entropy. 
Using the parametrization  
{
$q_i(\beta\varepsilon)=\exp(-\alpha-\beta\varepsilon_i)$ for the prior probabilities
in the variables  $\beta\varepsilon_i$, 
one obtains $\psi=H(p)-\alpha\sum_i p_i-\beta\sum_ip_i\varepsilon_i$.
Maximizing $\psi-\alpha''\sum_i p_i$ with Lagrange multiplier $\alpha''$ allows us to absorb the term $\alpha\sum_ip_i$ into the normalization constraint by identifying $\alpha'=\alpha+\alpha''$. 
Since the distribution $q$ is normalized, $\alpha=\alpha(\beta\varepsilon)$ 
is a function of $\beta\varepsilon$, and~one obtains the MC solution, 
 $\hat p=q(\beta\varepsilon)=\frac{1}{Z}\exp(-\beta \varepsilon_i)$. 
Here $Z=\exp(\alpha)$ is the normalization constant.
Defining~$\tilde H(\beta\varepsilon)=H(\hat p)$, we obtain in the maximum configuration that
$\hat\psi(\beta\varepsilon)=\tilde H-\alpha-\beta U$, 
with~$U=\sum_i \hat p_i\varepsilon_i$.
Moreover, $\hat\psi$ can be interpreted as the {\em Legendre transformation} of $H(p)-\alpha\sum_i p_i$, 
where~the term $-\sum_ip_i\beta\varepsilon_i$ transforms the variables $p_i$ into the adjoined variables, $\beta\varepsilon_i$.
Since~$P_{\rm multinomial}(N\hat p|q)$ asymptotically approaches $1$, we obtain $\hat \psi=0$  and therefore 
$U-\tilde H/\beta=-\alpha/\beta$, where $F=-\alpha/\beta$ is the {\em Helmholtz free energy}.
}
Since  $\tilde H(\beta\varepsilon)$ represents thermodynamic entropy in equilibrium thermodynamics, 
one might ask whether a similar transformation exists for driven systems. 

Let us parametrize the weights $q_i$ and $\lambda_i$ as
\begin{equation}
\begin{array}{lcl}
q_i&=&
{
\exp(-\alpha-\beta \varepsilon_i)
}
\\
1-\lambda_i&=&\exp(-\beta^* \varepsilon^*_i)\quad
\end{array}.
\end{equation} 

Note that $1-\lambda_i$ need not sum to $1$, so $\exp(-\beta^* \varepsilon^*_i)$ does not need a normalization factor. 
{
Moreover, from $\lambda_1=0$, it follows that $\varepsilon^*_1=0$.
}
Inserting these definitions in the cross-entropy in Equation~(\ref{crossent}), we obtain 
\begin{equation}
S_{\rm cross}(p^*,p,\bar \ell|q,\lambda)=S_{\rm D}+\bar\ell\alpha\sum_{i=1}^W p_i 
+ \beta^*\sum_{i=1}^W p^*_i\varepsilon^*_i+ \bar\ell\beta\sum_{i=1}^W p_i\varepsilon_i \quad, 
\end{equation} 
with---what we call---the \textit{driving entropy} 
\begin{equation}
S_{\rm D}=-\sum_{j=2}^W \left(\bar\ell p_j- p^*_j\right)\Delta s_j \quad.
\end{equation} 

Here $\Delta s_j=\log \lambda_j-\log Q_{j-1}$, which, in the given parametrization, 
is a function of $\beta^*\varepsilon^*$ and $\bar\ell\beta\varepsilon$.
In~the maximum  configuration, the terms $\bar\ell\beta\sum_i p_i\varepsilon_i$  and $\beta^*\sum_i p^*_i\varepsilon^*_i$ again 
are the Legendre transformation from variables $p$ and $p^*$ to the adjoined variables, $\hat\ell\beta\varepsilon$ 
and $\beta^*\varepsilon^*$. $\hat\ell$ is the MC of $\bar\ell$ and \mbox{$\psi= S_{\rm MC}- S_{\rm cross}$}.
By defining now the ``thermodynamic entropy,''
\mbox{$\tilde S_{\rm MC}(\hat\ell\beta\varepsilon,\beta^*\varepsilon^*)=\hat S_{\rm MC}/\hat\ell$}, where~
\mbox{$\hat S_{\rm MC}=S_{\rm MC}(\hat p,\hat p^*,\hat\ell)$}, and, analogously,  
 the ``thermodynamic driving entropy,'' \mbox{$\tilde S_{\rm D}(\hat\ell\beta\varepsilon,\beta^*\varepsilon^*)=
\hat S_{\rm D}/\hat\ell$}, one obtains
$\hat\psi(\hat\ell\beta\varepsilon,\beta^*\varepsilon^*)/\hat \ell=\tilde S_{\rm MC}-\tilde S_{\rm D}-\alpha
- \beta^*U^*/\hat\ell- \beta U$.  
Here~\mbox{$U^*=\sum_{i=1}^W \hat p^*_i\varepsilon^*_i$}, and $U=\sum_{i=1}^W \hat p_i\varepsilon^*_i$ is the ``internal energy.''
Since asymptotically (for large $N$) we have $\hat \psi=0$ again, it follows that
\begin{equation}
	U-\tilde S_{\rm MC}/\beta=-\alpha/\beta-\beta^*U^*/\beta-\tilde S_{\rm D}/\beta \quad.
\end{equation}

In  contrast to the maximum entropy principle for i.i.d. processes, where the inverse temperature  
and the normalization term of the distribution function are one-to-one related to the Lagrangian multipliers, 
this is no longer true for driven systems. 
The parameters $\alpha$, $\beta$, and $\beta^*$ are entangled in $S_{\rm D}$! 
{
However, this fact does not---as one might expect---destroy the overall Legendre structure
of the statistical theory of driven processes. It has, however,  a important consequence:
}
we can no longer expect notions such as the 
\textit{Helmholtz free energy} ({which has a clear interpretation for reversible processes}) to 
have the same meaning for driven systems.
It~is, however, exactly this kind of question  that the theory that we outlined in this paper may help us tackle. 
{
For example, in a driven system, we have to decide whether we call $U-\tilde S_{\rm MC}/\beta$ the Helmholtz free energy, $F$, 
or if it is $-\alpha/\beta$. 
If we identify $F=-\alpha/\beta$ and define the ``{\em driven} Helmholtz free energy'' as, $F_{\rm D}=U-\tilde S_{\rm MC}/\beta$, then we obtain 
$F_{\rm D}=F-\beta^*U^*/\beta-\tilde S_{\rm D}/\beta$.
This means that, for driven systems, it remains possible to switch between thermodynamic potentials using 
Legendre transformations, e.g., by switching from $U$ to $F_{\rm D}$. 
Most importantly, the landscape of thermodynamic potentials is now enriched with two additional terms, 
$W_{\rm D}= \beta^*U^*/\beta$  and $Q_{\rm D}= \tilde S_{\rm D}/\beta$, 
which incorporate the driving process and its interactions with the system, 
\begin{equation}
	U-\tilde S_{\rm MC}/\beta=F-W_{\rm D}-Q_{\rm D} \quad.
\end{equation}

It is tempting to interpret $W_{\rm D}$ and $Q_{\rm D}$ as irreversible work and dissipated heat, respectively. 
However, a clear interpretation of those terms can only be achieved within the concrete context of a given application. }

\section{Conclusions}
We have shown that it is possible to derive a consistent \textit{maximum configuration} framework for driven dissipative systems  
that  
are a combination of a state-dependent driving process and a relaxation process that can always can be modeled as an SSR process. 
The considered processes must  exhibit stationary distributions. 
The presented framework is a statistical theory that not only allows us to compute visiting distribution functions $p$ from an appropriate entropy functional but also allows us to predict properties, such as its state-dependent driving rates, $p^*$, or  
the average relaxation times, $\bar \ell$. 
We find that the visiting distributions are in exact agreement with alternatively computable analytical solutions.   

Remarkably, the maximum configuration entropy functional that emerges within the context of driven processes 
decouples completely from the detailed constraints posed by the interactions with the driving system. 
In this sense, it represents a property that is universally shared by processes within the considered class of driven processes. 
As a consequence, the maximum configuration mantra: maximize~\textit{entropy} $+$  \textit{constraint terms} to  
predict the maximum configuration, remains valid also for driven systems. 
One only needs to keep in mind that the ``reasonable''  constraint terms that specify the interactions of the system with the driving process may have a more complicated structure than we are used to from i.i.d. processes.
The presented theory further allows us to compute a large number of expectation values that correspond to 
physically observable macro variables of the system. 
This~includes the possibility of studying fluctuations of the system and the driving process together in the same framework.

The structure of the maximum configuration entropy and the corresponding maximum entropy principle is interesting by itself. 
By introducing the relative frequencies $p^*$ of the driving process and the average free relaxation time, $\bar \ell$, 
as independent quantities, the entropy decouples from the actual driving process and from how the system is linked to it.  
At the same time, this is only possible when the distributions $p^*$ and $p$ become functionally entangled.  
$\bar \ell$ effectively quantifies the strength of this entanglement. 
For slowly driven systems, it is the ground state that is entangled with all other states. 
For fast driving, all states become ``hierarchically''  entangled.
Remarkably, however complicated this entanglement may be, we still can compute all of the involved distribution functions.
As for i.i.d. processes, the maximum configuration asymptotically (for large systems) dominates the system behavior. 
Therefore, the maximum configuration entropy, at its maximum configuration (determined by the macro state of the driven system), 
can be given a thermodynamic meaning.

Finally, we discussed the thermodynamic structure of the presented statistical theory. 
The type of processes we can consistently describe with the presented framework include (kinetic) energy dissipations 
in elastic collision processes (see \cite{Ref:SSR_ElastColl}). 
It is therefore not only conceivable, but in fact highly probable, that for a wide class of driven processes, 
a consistent thermodynamics can be derived from 
{
the statistical theory. Despite the more complicated relations between distribution functions and system parameters, 
the Legendre structure of equilibrium thermodynamics survives in the statistical theory of the driven processes presented here. 
The landscape of macro variables ($U$, $U^*$, $\beta$, $\beta^*$, \dots) 
and potentials ($F$, $F_{\rm D}$, $Q_{\rm D}$, \dots) becomes richer, and 
now includes terms that describe the driving process and how it couples to the relaxing system. 
Most importantly, however, the~underlying statistical theory allows us to analyze   
all the differential relations between macro variables and potentials that we might consider. 
These may depend on the parametrization of $q$ and $\lambda$. 
The~presented work makes it conceivable that statistical theories can be constructed for situations, 
where multiple, not just two, driving and relaxing processes are interacting with one another.  
This results in more complicated functional expressions for  
MC entropy, cross-entropy, additional macro state variables, and thermodynamic potentials 
but does not introduce fundamental or conceptual problems.
}

RH and ST conceptualized the work and wrote the paper.
We thank B. Corominas-Murtra for helpful discussions and acknowledge support from the Austrian Science Foundation 
FWF under P29032.
 
\bibliographystyle{mdpi}

\renewcommand\bibname{References}

\end{document}